\documentstyle[12pt]{article}
\begin{document}

\title{Path integral in the simplest Regge calculus model}
\author{V.M.Khatsymovsky \\
 {\em Budker Institute of Nuclear Physics} \\ {\em
 Novosibirsk, 630090,
 Russia} \\ {\em E-mail address: khatsym@inp.nsk.su}}
\date{}
\maketitle
\begin{abstract}
The simplest (3+1)D Regge calculus model (with
three-dimensional discrete space and continuous time) is
considered which describes evolution of the simplest closed
two-tetrahedron piecewise flat manifold in the continuous
time. The measure in the path integral which describes
canonical quantisation of the model in terms of area
bivectors and connections as independent variables is
found. It is shown that selfdual-antiselfdual splitting of
the variables simplifies the integral although does not
admit complete separation of (anti-)selfdual sector.
\end{abstract}
\newpage
Path integral is probably the most efficient way to
quantisation of the discrete theory. In particular, this
approach is widely used in Regge calculus, including the
physical 4D case \cite{Ham,Sil}. The path integral measure
is usually chosen as the simplest among the invariant ones.
Another approach is based on constructing state sum
invariants of the 4D Regge manifolds
\cite{Car,Wil,Bar} in analogy with how it is done in the 3D
case where such invariants being functionals of only 2D
triangulated boundary of such manifold exist and do not
depend on the refinement of the interior tetrahedrons
\cite{Pon,Has,Tur,Bar1,Bar2}. Thus, quantum measure is
defined there by topological requirements.

If description of the system is known in the canonical
form, the most natural choice of path integral measure is
that based on the canonical quantisation. Knowing canonical
form of the theory requires that action for the system in
the continuous time be known. In the given paper we
consider path integral just for the simplest 3D Regge
manifold evolving in the continuous time. Canonical form of
this model as limiting case of the completely discrete 4D
Regge manifold has been considered in \cite{Kha} in the
area bivector and connection variables on the base of the
canonical formulation of the general Regge manifold
\cite{Kha1}. The 3D Regge
manifold consists of two similar tetrahedrons with mutually
identified vertices. Remarkable feature of the related
continuous time Regge calculus action is it's notational
analogy and correspondence of constraints with the case of
the continuum general relativity (GR) action in the
Hilbert-Palatini (HP) form, whereas the number of the
degrees of freedom is finite, and "large distances" (i.e.
those considerably larger than the size of tetrahedron) are
absent. Therefore the path integral for this system can be
considered as that modelling path integral for the
continuum GR on condition that long-distance effects
(gravity waves) are excluded.  The typical feature of the
path integral measure obtained is that it vanishes for
symmetrical configurations of the system. This indicates
that quantum fluctuations at these points are definitive.

Also path integral is convenient to handle the problem of
implementing the so-called Ashtekar variables \cite{Ash} in
the Regge calculus (see review paper \cite{Imm}) on quantum
level. From general viewpoint, there are the two possible
approaches to this problem: to put Ashtekar theory on the
Regge lattice or to formulate Regge calculus in the
Ashtekar-like variables. The recent approaches \cite{Imm}
seem to be of the first type, we try in our work the second
one via splitting the antisymmetric tensor variables into
the selfdual and antiselfdual parts. It turns out that
although there is no complete separation of selfdual and
antiselfdual sectors as it occurs in the continuum GR, the
considered procedure simplifies path integral considerably.

We start with the canonical structure of our model
\cite{Kha}. The Lagrangian with all the constraints added
reads:
\begin{eqnarray}                                         %1
L = \sum_{\mu}{\pi^\mu\circ\bar{\Omega}_\mu
\dot{\Omega}_\mu} & + & \sum_{\mu > \nu}{
\dot{\psi}_{\mu\nu}\pi^\mu\circ\pi^\nu}
- \sum_{\mu}v_\mu H_\mu - h\circ\sum_{\mu}{\pi^\mu} -
\tilde{h}\circ\sum_{\mu}{
\Omega_\mu\pi^\mu\bar{\Omega}_\mu}\nonumber\\
& + & \sum_{\mu > \nu}{\psi_{\mu\nu}\Delta
(\pi^\mu\circ\pi^\nu )} +
\sum_{\mu}{\kappa_\mu\pi^\mu*\pi^\mu} + \sum_{\mu >
\nu}{\kappa_{\mu\nu}\pi^\mu*\pi^\nu}.
\end{eqnarray}
Here $\mu$, $\nu$, $\lambda$, $\rho$,... = 0, 1, 2, 3 run
over four vertices of tetrahedrons; $\pi^\mu_{ab}$ =
$-\pi^\mu_{ba}$ are bivectors of the triangles opposite (in
3D leaf) to the vertices $\mu$; $a$, $b$, $c$, $d$,... = 0,
1, 2, 3 are local frame SO(3,1) indices (SO(4) in the case
of Euclidean signature); $\Omega^{ab}_\mu$ $\subset$
SO(3,1); $A\circ B$ $\equiv$ $(1/2){\rm tr}A\bar{B}$ for
the bivectors $A^{ab}$ and $B^{ab}$; $\,^*\! A_{ab}$
$\equiv$ $(1/2)\varepsilon_{abcd}A^{cd}$; $A*B$ $\equiv$
$A\circ (\,^*\! B)$; $h^{ab}$, $\tilde{h}^{ab}$ are
multipliers at Gaussian constraints; the scalars $v_\mu$
enter also $\Delta$ $\equiv$ $\sum_{\mu}{v_\mu\Delta_\mu}$
and are, therefore, multipliers at the Hamiltonian
constraints which are not $H_\mu$ but ${\cal H}_\mu$
$\equiv$ $H_\mu$ - $\Delta_\mu\sum_{\nu > \lambda}
{\psi_{\nu\lambda}\pi^\nu\circ\pi^\lambda}$ and
\begin{eqnarray}                                         %2
\Delta_\mu\pi^\nu & \equiv & \sum_{\lambda ,\rho}{
\varepsilon_{\mu\nu\lambda\rho}n_{\mu\lambda}},\\
H_\mu & = & \sum_{\nu}{|n_{\mu\nu}\!|\arcsin{            %3
\frac{n_{\mu\nu}\circ R_{\mu\nu}}{|n_{\mu\nu}\!|}}
+ \varphi^\nu_\mu\pi^\nu\circ\Delta_\mu\pi^\nu},\\
n_{\mu\nu} & = & \sum_{\lambda ,\rho}{                   %4
\varepsilon_{\mu\nu\lambda\rho}\left (u^\lambda_\mu\pi^\rho
+{1\over 2}[\pi^\lambda ,\pi^\rho]\right )},\\
R_{\mu\nu} & = & {1\over 2}\sum_{\lambda ,\rho\neq\mu ,\nu}
 {\left(\bar{\Omega}_{\mu\rho}                           %5
\Omega_{\mu\lambda}
\right )^{\varepsilon_{\mu\nu\lambda\rho}}},\\
\label{exp-pi}
\Omega_{\mu\nu} & = & \Omega_\nu\Pi^\nu_\mu,\\           %6
\Pi^\nu_\mu & \equiv & \exp{(\varphi^\nu_\mu             %7
\pi^\nu + \,^*\!\varphi^\nu_\mu\,^*\!\pi^\nu)}.
\end{eqnarray}
The completely antisymmetric symbol is defined so that
$\varepsilon_{0123}$ = +1.
The $u^\nu_\mu$, $\varphi^\nu_\mu$, $\,^*\!\varphi^\nu_\mu$
are nondynamical independent variables (not appearing in
the kinetic term). These enter Lagrangian nonlinearly. The
$\psi_{\mu\nu}$, $\kappa_\mu$ and $\kappa_{\mu\nu}$ are
variables entering $L$ linearly of which $\psi_{\mu\nu}$
are dynamical ones. Of the six variables $\psi_{\mu\nu}$
(or $\kappa_{\mu\nu}$) only two are independent ones, and
in \cite{Kha} the $\psi_{12}$, $\psi_{23}$ (and
$\kappa_{12}$, $\kappa_{23}$) were chosen as the only
nonzero ones. Now, in order to formulate the theory in the
form explicitly symmetrical w.r.t. the permutation of the
vertices 0, 1, 2, 3 we assume that all $\psi_{\mu\nu}$,
$\kappa_{\mu\nu}$ are not identically zero, but are subject
to the following four relations on $\psi_{\mu\nu}$:
\begin{equation}
\Psi_\mu \equiv \sum_{\nu ,\lambda\neq\mu}{              %8
\psi_{\nu\lambda}} = 0,
\end{equation}
and four ones on $\kappa_{\mu\nu}$:
\begin{equation}
\sum_{\nu ,\lambda\neq\mu}{                              %9
\kappa_{\nu\lambda}} = 0.
\end{equation}
Let us take into account these relations on
$\kappa_{\mu\nu}$ with the help of Lagrange multipliers
$C_\mu$. Then we can consider $\kappa_{\mu\nu}$ as six
independent variables, coefficients at the following
constraints:
\begin{equation}
Z_{\mu\nu} \equiv \pi^\mu*\pi^\nu                       %10
+ \sum_{\lambda\neq\mu ,\nu}{C_\lambda} = 0.
\end{equation}
The $C_\mu$ are considered then as independent variables.

Now passing to the Hamiltonian formalism we need the
extended phase space coordinatised by the following
canonical pairs $(p,q)$: $(\pi^\mu ,\Omega_\mu)$,
$(\tilde{\psi}^{\mu\nu},\psi_{\mu\nu})$,
$(\tilde{C}^\mu ,C_\mu)$. The full system of constraints
$\Phi$ consists of the first class $\eta$ and second class
$\Theta$ ones. The coefficients at $\eta$ in $L$ are $h$,
$\tilde{h}$ and $\kappa_\mu$. The $\Theta$ consists of
${\cal H}_\mu$, $\Psi_\mu$, $Z_{\mu\nu}$ and the following
constraints on the newly defined momenta:
\begin{equation}
\tilde{C}^\mu = 0                                       %11
\end{equation}
and
\begin{equation}
\tilde{\Psi}^{\mu\nu} \equiv \tilde{\psi}^{\mu\nu}      %12
- \pi^\mu\circ\pi^\nu = 0.
\end{equation}

Next we introduce the gauge conditions $\chi$ = 0 by the
number of the first class constraints $\eta$ such that
${\rm Det}\{\eta ,\chi\}$ $\neq$ 0. Then the path integral
measure takes the standard form:
\begin{eqnarray}
d\mu_\chi & = & e^{i\int{Ldt}}\delta (\Phi_\chi)
{\rm Det}^{1/2}\{\Phi_\chi ,\Phi_\chi\}DpDq
\nonumber\\
\label{d-mu-chi}
& = & e^{i\int{Ldt}}{\rm Det}^{1/2}\{\Theta ,\Theta\}   %13
\delta (\Theta )\delta (\eta )\delta (\chi ){\rm Det}
\{\eta ,\chi\}DpDq
\end{eqnarray}
where $\Phi_\chi$ = $(\Phi ,\chi)$. For the specific form
of the kinetic term for $(p,q)$ = $(\pi ,\Omega)$ one gets
$DpDq$ = $d^6\pi{\cal D}^6\Omega$, ${\cal D}^6\Omega$ being
the Haar measure on SO(3,1), and under the Poisson brackets
(PB) one should imply the expression
\begin{equation}
\label{{}}
\{ f,g\} = \pi\circ [f_\pi ,g_\pi ]                     %14
+ f_\pi\circ\bar{\Omega}g_\Omega
- g_\pi\circ\bar{\Omega}f_\Omega
\end{equation}
(for the functions of purely $\pi$, $\Omega$; taking into
account dependence on other variables is straightforward).
In \cite{Kha,Kha1} this form was used as that describing
Hamiltonian dynamics. Although quite expectable, this can
be proved by considering standard kinetic term
$(1/2){\rm tr}\bar{P}\dot{\Omega}$ in the further extended
phase space of canonical pairs $(P,\Omega)$,
$(\tilde{\psi},\psi)$, $(\tilde{C},C)$ and substituting
into (\ref{d-mu-chi}) instead of $\Phi_\chi$ the set
$\Phi^\prime_\chi$ = $(\Phi_\chi ,\tau)$ extended by
inclusion the constraints $\tau$ specifying the 4 $\times$
4 matrices $P$, $\Omega$ to have the form $\Omega$ $\in$
SO(3,1), $P$ = $\Omega\pi$, $\bar{\pi}$ = $-\pi$. Then
${\rm Det}\{\Phi^\prime_\chi ,\Phi^\prime_\chi\}$ =
${\rm Det}\{\Phi_\chi ,\Phi_\chi\}_{{\rm D}(\tau )}$ where
$\{\cdot ,\cdot\}_{{\rm D}(\tau )}$ are Dirac brackets (DB)
w.r.t. the set $\tau$. The
 $\{\cdot ,\cdot\}_{{\rm D}(\tau )}$ turn out to coincide
just with the earlier defined brackets (\ref{{}}). The Haar
measure just arises when integrating over $DPD\Omega$ the
newly arising $\delta$-functions of $\tau$,
\begin{equation}
\delta (\tau ) = \prod_{\mu}{\delta (\bar{\Omega}_\mu   %15
\Omega_\mu - 1)\delta (\bar{\Omega}_\mu P^\mu
+ \bar{P}^\mu\Omega_\mu)}.
\end{equation}

Now consider an important consequence of finiteness (in the
Euclidean case) of the volume of the gauge group. The gauge
transformations are some rotations changing $\Omega_\mu$,
$\,^*\!\varphi^\mu_\nu$. At the same time, the gauge fixing
factor $\delta (\chi){\rm Det}\{\eta ,\chi\}$ in the path
integral measure just arises upon dividing by the volume of
the group of gauge transformations (generated by I class
constraints $\eta$ via PB $\{\eta ,\cdot\}$). Therefore, if
we consider Euclidean case (or Lorentzian one in the sense
of analytical continuation from the Euclidean case) we can
insert this volume back into measure simply omitting gauge
fixing factor without a risk of getting a new infinity. The
measure takes the form
\begin{equation}
d\mu =  e^{i\int{Ldt}}{\rm Det}^{1/2}\{\Theta ,\Theta\} %16
\delta (\Theta )\delta (\eta )DpDq.
\end{equation}

It is convenient to divide the set of second class
constraints $\Theta$ into two second class subsets
$\vartheta$ and $\iota$ so that
${\rm Det}\{\Theta ,\Theta\}$ =
${\rm Det}\{\vartheta ,\vartheta\}_{{\rm D}(\iota )}$. Let
the set $\vartheta$ consists of the four Hamiltonian
constraints ${\cal H}_\mu$, the $\iota$ be the set of other
second class constraints $\Psi_\mu$, $Z_{\mu\nu}$,
$\tilde{C}^\mu$, $\tilde{\Psi}^{\mu\nu}$. To calculate the
DB $\{{\cal H}_\mu ,{\cal H}_\nu\}_{{\rm D}(\iota )}$ let
us first define the projection ${\cal H}^\perp_\mu$ of
${\cal H}_\mu$ onto the subspace orthogonal w.r.t. the PB
to the space of differentials of the set $\iota$:
\begin{equation}
{\cal H}^\perp_\mu = {\cal H}_\mu - \sum_{\lambda > \nu}{
A^{\lambda\nu}_\mu Z_{\lambda\nu}} - \sum_{\nu}{B_{\mu\nu}
\tilde{C}^\nu} - \sum_{\nu}{E^\nu_\mu\Psi_\mu}
- \sum_{\lambda > \nu}{
F_{\mu\lambda\nu}\tilde{\Psi}^{\lambda\nu}}             %17
\end{equation}
where the coefficients $A$, $B$, $E$, $F$ are defined so
that $\{{\cal H}^\perp_\mu , \iota\}$ = 0. For example,
commuting ${\cal H}^\perp_\mu$ with constraints
$Z_{\lambda\nu}$ and $\Psi_\lambda$ gives, respectively,
\begin{equation}
\label{B}
- \sum_{\rho \neq \lambda , \nu}{B_{\mu\rho}}           %18
+ \left\{\sum_{\rho > \epsilon}{F_{\mu\rho\epsilon}
\pi^\rho\circ\pi^\epsilon}, \pi^\lambda*\pi^\nu\right\}
+ \{{\cal H}_\mu , \pi^\lambda*\pi^\nu\} = 0
\end{equation}
and
\begin{equation}
\label{F}
\sum_{\nu , \rho \neq \lambda}{F_{\mu\nu\rho}} = 0.     %19
\end{equation}
These eqs. allow to find $B_{\mu\nu}$ and
$F_{\mu\nu\lambda}$. Commutation with $\tilde{C}^\nu$ and
$\tilde{\Psi}^{\lambda\nu}$ allows to find
$A^{\lambda\nu}_\mu$ and $E^\nu_\mu$ from the analogous
system. In the calculations of such kind commutators
between different $\pi$-bilinears, $\pi^\mu\circ\pi^\nu$
and $\pi^\mu*\pi^\nu$, arise. These commutators by
(\ref{{}}) reduce to the following trilinears:
\begin{eqnarray}
\pi^\mu\circ [\pi^\nu , \pi^\lambda ]& = &
\pi^\mu * [\pi^\nu ,\,^*\!\pi^\lambda ] = 0,\nonumber\\
\pi^\mu * [\pi^\nu , \pi^\lambda ] & = & - V^2\sum_{\rho}{
\varepsilon_{\rho\mu\nu\lambda}},                       %20
\end{eqnarray}
$V$ being the volume of the parallelepiped spanned by the
links issuing from any tetrahedron vertex. These equalities
easily follow from explicit expressions of bivectors in
terms of link vectors. The consequence is that commutators
of such bilinears of the type of scalar-scalar and
pseudoscalar-pseudoscalar vanish, only those of the
scalar-pseudoscalar type are nonzero. For example,
\begin{equation}
\left\{\sum_{\nu > \lambda}{F_{\mu\nu\lambda}           %21
\pi^\nu\circ\pi^\lambda}, \pi^0*\pi^1\right\}
= V^2(F_{12} + F_{03} - F_{31} - F_{02}).
\end{equation}
Finding $A$, $B$, $E$, $F$ we can find the DB of interest,
\begin{eqnarray}
\{{\cal H}_\mu , {\cal H}_\nu\}_{{\rm D}(\iota )} & \equiv
 & \{{\cal H}^\perp_\mu , {\cal H}^\perp_\nu\} =
\{H_\mu , H_\nu\} + \sum_{\lambda > \rho}{\psi
_{\lambda\rho}[\Delta_\nu , \Delta_\mu]\pi^\lambda
\circ\pi^\rho}\nonumber\\
& - & \sum_{\stackrel{\lambda > \rho}{
\sigma > \phi}}{\{H_\nu , \pi^\lambda*\pi^\rho\}
K_{\lambda\rho\sigma\phi}(\Delta^\Omega_\mu - \Delta_\mu)
(\pi^\sigma\circ\pi^\phi)}\nonumber\\
& + & \sum_{\stackrel{\lambda > \rho}{
\sigma > \phi}}{\{H_\mu , \pi^\lambda*\pi^\rho\}
K_{\lambda\rho\sigma\phi}(\Delta^\Omega_\nu - \Delta_\nu)
(\pi^\sigma\circ\pi^\phi)}                              %22
\end{eqnarray}
where we have introduced the notation $\Delta^\Omega_\mu
(\cdot )$ $\equiv$ $\{H_\mu , \cdot\}$. The reason for such
notation is that for the particular case of the solution to
the eqs. of motion for $\Omega$ when $R_{\mu\nu}n_{\mu\nu}
\bar{R}_{\mu\nu}$ = $n_{\mu\nu}$ for each $\mu\nu$ (i.e.
when $\Omega$ is a true rotation between the local frames)
the operator $\Delta^\Omega_\mu$ coincides with
$\Delta_\mu$ (variation at shifting the point $\mu$ to the
next time leaf). The $K_{\mu\nu\lambda\rho}$ can be called
an inversed to the matrix $\{\pi^\mu\circ\pi^\nu ,
\pi^\lambda * \pi^\rho\}$,
\begin{equation}
K_{\mu\nu\lambda\rho} = \frac{1}{12V^2}\sum_{\epsilon}{
(\varepsilon_{\lambda\rho\nu\epsilon}\delta^\rho_\mu    %23
 + \varepsilon_{\lambda\rho\mu\epsilon}\delta^\rho_\nu
 - \varepsilon_{\lambda\rho\nu\epsilon}\delta^\lambda_\mu
- \varepsilon_{\lambda\rho\mu\epsilon}\delta^\lambda_\nu)}.
\end{equation}
(The matrix $\{\pi^\mu\circ\pi^\nu , \pi^\lambda *
\pi^\rho\}$ is degenerate, and a strict sense to this term
is given by solving the eqs. (\ref{B}), (\ref{F}).) These
DB can be further transformed using the following two
facts. First, note that the operation
\begin{equation}
f \rightarrow g = f - \sum_{\stackrel{\mu > \nu}        %24
{\lambda > \rho}}{\{f , \pi^\mu * \pi^\nu\}
K_{\mu\nu\lambda\rho}\pi^\lambda\circ\pi^\rho}
\end{equation}
on the space of linear combinations of scalars
$\pi^\mu\circ\pi^\nu$ projects onto the subspace of such
combinations of only the squares $|\pi^\mu|^2$. On these
squares $\Delta^\Omega_\mu$ = $\Delta_\mu$ (as consequence
of the eqs. of motion for $\varphi^\mu_\nu$). This allows
to simplify contribution to $\{H_\mu , g\}$ which results
from the dependence of $f$ on the scalars
$\pi^\mu\circ\pi^\nu$. Second, there is some complication
connected with the (exponential) dependence of
$\Omega_{\mu\nu}$ on $\pi^\nu$ (\ref{exp-pi}). Suppressing
indices, consider a function $f(\pi , \tilde{\Omega})$ of
$\pi$ and $\tilde{\Omega}$ = $\Omega\Pi$, $\Pi$ =
$\exp{(\varphi\pi + \,^*\!\varphi\,^*\!\pi)}$. The
following formula can be obtained:
\begin{equation}
\left (\frac{\partial f}{\partial\pi}\right )_\Omega =  %25
\left (\frac{\partial f}{\partial\pi}\right )
_{\tilde{\Omega}} - {1\over |\pi|^2}\left [\pi ,
\Pi\bar{\tilde{\Omega}}\frac{\partial f}{\partial\tilde{\Omega}}
\bar{\Pi} - \bar{\tilde{\Omega}}\frac{\partial f}
{\partial\tilde{\Omega}}\right ],
\end{equation}
where both derivatives $\partial/\partial\pi$ and
$\bar{\tilde{\Omega}}\partial/\partial\tilde{\Omega}$ are
implied to be antisymmetrised and subscript at $\partial f/
\partial\pi$ denotes the quantity considered as constant at
the differentiating. This allows to simplify expression for
the PB of the two functions $f(\pi , \Omega_1)$ and
$g(\pi ,\Omega_2)$ where $\Omega_i$ = $\Omega\Pi_i$,
$\Pi_i$ = $\exp{(\varphi_i\pi
+ \,^*\!\varphi_i\,^*\!\pi)}$:
\begin{equation}
\{f , g\} = \{f , g\}_\Pi + \frac{1}{|\pi|^2}\pi\circ   %26
\left (\left [\bar{\Omega_1}{\partial f\over\partial
\Omega_1},\bar{\Omega_2}{\partial g\over\partial\Omega_2}
\right ] - \left [\Pi_1\bar{\Omega_1}{\partial f\over
\partial\Omega_1}\bar{\Pi}_1,\Pi_2\bar{\Omega_2}{\partial g
\over\partial\Omega_2}
\bar{\Pi}_2\right ]\right ).
\end{equation}
Here $\{f , g\}_\Pi$ are "naive" PB taken in the formal
assumption that $\Pi_i$ are constants. The second term in
this formula is responsible for the dependence of $\Pi_i$
on $\pi$. It vanishes at $\Pi_1$ = $\Pi_2$ as it should
because exponential can be eliminated by redefining
$\Omega$ in this case.

To represent the DB of the Hamiltonian constraints of
interest in the form resembling that of commutators in the
continuum GR (in the HP form) we introduce the notation
\begin{equation}
\label{h}
h_\mu = \sum_{\nu}{{n_{\mu\nu}\circ R_{\mu\nu}\over     %27
\cos{\alpha_{\mu\nu}}}}
\end{equation}
where $\alpha_{\mu\nu}$ is angle defect,
$\sin{\alpha_{\mu\nu}}$ = $n_{\mu\nu}\circ R_{\mu\nu}/
|n_{\mu\nu}|$. The $h_\mu$ arises when we differentiate
${\cal H}_\mu$ as complex function over phase variables on
which it depends through the angle defects. Note also that
the dependence of ${\cal H}_\mu$ on phase variables through
the nondynamical
ones $u^\nu_\mu$, $\varphi^\nu_\mu$,
$\,^*\!\varphi^\nu_\mu$ need not be taken into account at
the calculating the first derivatives due to the eqs. of
motion for the nondynamical variables. The DB of interest
read
\begin{eqnarray}
\{{\cal H}_\mu , {\cal H}_\nu\}_{{\rm D}(\iota )} & = &
\{h_\mu , h_\nu\}_{\alpha , \Pi} + \sum_{\lambda}{\frac{1}
{|\pi^\lambda|^2}\pi^\lambda\circ\left (\left [\bar{\Omega}
_{\lambda\mu}{\partial h_\mu\over\partial\Omega
_{\lambda\mu}},\bar{\Omega}_{\lambda\nu}{\partial h_\nu
\over\partial\Omega_{\lambda\nu}}\right ]\right.}
\nonumber\\
 & - & \left.\left [
\Pi^\lambda_\mu\bar{\Omega}_{\lambda\mu}{\partial h_\mu
\over\partial\Omega_{\lambda\mu}}\bar{\Pi}^\lambda_\mu ,
\Pi^\lambda_\nu\bar{\Omega}_{\lambda\nu}{\partial h_\nu
\over\partial\Omega_{\lambda\nu}}\bar{\Pi}^\lambda_\nu
\right ]\right )\nonumber\\
& + & \sum_{\lambda}{[(\alpha_{\nu\lambda} - \tan\alpha
_{\nu\lambda})\Delta_\mu |n_{\nu\lambda}|
- (\alpha_{\mu\lambda} - \tan\alpha
_{\mu\lambda})\Delta_\nu |n_{\mu\lambda}|]}\nonumber\\
& + & {1\over 2}\sum_{\lambda}{(\varphi^\lambda_\nu\Delta
_\mu\Delta_\nu - \varphi^\lambda_\mu\Delta_\nu\Delta_\mu)
(\pi^\lambda\circ\pi^\lambda)}
- \sum_{\lambda > \rho}{\psi_{\lambda\rho}
[\Delta_\mu , \Delta_\nu](\pi^\lambda\circ\pi^\rho)}
\nonumber\\
& + & \sum_{\stackrel{\lambda > \rho}{\sigma > \phi}}{
\{h_\mu ,\pi^\lambda*\pi^\rho\}_{\alpha , \Pi}
K_{\lambda\rho\sigma\phi}[\{h_\nu ,
\pi^\sigma\circ\pi^\phi\}_{\alpha , \Pi} - \Delta_\nu
(\pi^\sigma\circ\pi^\phi )]}\nonumber\\
& - & \sum_{\stackrel{\lambda > \rho}{\sigma > \phi}}{
\{h_\nu ,\pi^\lambda*\pi^\rho\}_{\alpha , \Pi}
K_{\lambda\rho\sigma\phi}[\{h_\mu ,
\pi^\sigma\circ\pi^\phi\}_{\alpha , \Pi} - \Delta_\mu
(\pi^\sigma\circ\pi^\phi )]}                            %28
\end{eqnarray}
where $\Pi^\nu_\mu$ = $\exp{(\varphi^\nu_\mu\pi^\nu +
\,^*\!\varphi^\nu_\mu\,^*\!\pi^\nu )}$, and the subscript
$\alpha$ at the PB means that defect angles in the
denominator in the definition of $h_\mu$ (\ref{h}) are
considered formally as constants. The first term in the
formula obtained is notationally close analog of the PB
of the combinations of vector (with coefficients
$u^\nu_\mu$) and Hamiltonian constraints in the continuum
GR.

The DB obtained define then the Jacobian factor entering
path integral:
\begin{eqnarray}
{\rm Det}^{1/2}\{\Theta , \Theta\} & = & |\{{\cal H}_0 ,
{\cal H}_1\}_{{\rm D}(\iota )}\{{\cal H}_2 ,
{\cal H}_3\}_{{\rm D}(\iota )} + \{{\cal H}_0 ,
{\cal H}_2\}_{{\rm D}(\iota )}\{{\cal H}_3 ,
{\cal H}_1\}_{{\rm D}(\iota )}\nonumber\\
 & + & \{{\cal H}_0 ,
{\cal H}_3\}_{{\rm D}(\iota )}\{{\cal H}_1 ,
{\cal H}_2\}_{{\rm D}(\iota )}|.                        %29
\end{eqnarray}
The rest factors follow by integrating $\delta$-functions
in $d\mu$. Integration over $d^4\tilde{C}$,
$d^6\tilde{\psi}$ and $d^6\pi^0$ eliminates
$\delta^4(\tilde{C})$ $\delta^6(\tilde{\Psi})$
$\delta^6(\sum_{\mu}{\pi^\mu})$. Integration over $d^4C$
reduces $\delta^6(Z)$ to $\delta$'s of some two independent
constraints of the type of $\pi^\mu*\pi^\nu$, $\mu$ $\neq$
$\nu$. Together with $\delta^4(\{\pi^\mu*\pi^\mu\})$ this
gives $\delta^6(\{\pi^\alpha*\pi^\beta\})$ ($\alpha$,
$\beta$, $\gamma$ = 1, 2, 3). Finally, the $\delta^4(\psi)$
annihilate four integrations in $d^6\psi$, and we are left
with, e.g., $d^2\psi$ = $d\psi_{21}$ $d\psi_{32}$ while
$\psi_{10}$ = $\psi_{32}$, $\psi_{20}$ = $\psi_{31}$ =
-$\psi_{21}$ - $\psi_{32}$, $\psi_{30}$ = $\psi_{21}$. The
result reads
\begin{eqnarray}
d\mu & = & \exp{\left\{i\int{dt\left [\sum_{\alpha}
{\pi^\alpha\circ (\bar{\Omega}_\alpha\dot{\Omega}_\alpha
- \bar{\Omega}_0\dot{\Omega}_0)} + \sum_{\lambda > \nu}{
\dot{\psi}_{\lambda\nu}\pi^\lambda\circ\pi^\nu}\right ]}
\right\}}{\rm Det}^{1/2}\{\Theta , \Theta\}\nonumber\\
 & & \delta^6\left [\sum_{\alpha}{(\Omega_\alpha\pi^\alpha
\bar{\Omega}_\alpha - \Omega_0\pi^\alpha\bar{\Omega}_0)}
\right ]\prod_{\mu}{\delta\left [H_\mu - \sum_{\lambda >
\nu}{\psi_{\lambda\nu}\Delta_\mu (\pi^\lambda\circ\pi^\nu)}
\right ]}\nonumber\\
\label{d-mu}
 & & \prod_{\alpha\geq\beta}{\delta (\pi^\alpha*
\pi^\beta)}d^2\psi\prod_{\alpha}{d^6\pi^\alpha}
\prod_{\mu}{{\cal D}^6\Omega_\mu}.                      %30
\end{eqnarray}
Some two of the four $\delta ({\cal H}_\mu)$'s can, in
principle, be integrated over $d^2\psi$ and the variables
$\psi_{\mu\nu}$ be excluded as functions of $\pi$,
$\Omega$. Note that despite of that $\psi$ seem to enter
${\cal H}_\mu$ linearly, the implicit dependence on $\psi$
through $u$, $\varphi$, $\,^*\!\varphi$ is much more
complex.

Due to the Jacobian factor the measure vanishes for
symmetrical configurations of the system for which the DB
$\{{\cal H}_\mu , {\cal H}_\nu\}_{{\rm D}(\iota)}$ vanish.

The remaining $\delta$-functions of $\pi^\mu*\pi^\nu$
reflect the tetrad structure of bivectors, in particular,
possibility to place the tetrahedron into 3D space and
consider it's area vectors instead of bivectors. To pass to
the 3D vector formulation in the explicitly invariant way
one can choose splitting the bivectors and generators of
the connections into selfdual and antiselfdual parts. To
this end we expand
\begin{equation}
\pi^\mu_{ab} = \,^+\!\pi^\mu_{ab} + \,^-\!\pi^\mu_{ab},
~~~\,^{\pm}\!\pi^\mu_{ab} = \,^{\pm}\!\Sigma^i_{ab}
\,^{\pm}\!\pi^\mu_i/2                                   %31
\end{equation}
where $\,^{\pm}\!\Sigma^i_{ab}$ form basis of (anti-)
selfdual matrices obeying algebra of Pauli ones times
$\sqrt{-1}$. The $\,^{\pm}\!\vec{\pi}^\mu$ = $\{\,^{\pm}\!
\pi^\mu_i|i=1,2,3\}$ are vectors in an abstract complex
3D space. The generator $\omega_\mu$ of the rotation
$\Omega$ = $\exp{\omega_\mu}$ can be decomposed in the same
way into $\,^+\!\omega_\mu$ and $\,^-\!\omega_\mu$ which
act on the abstract space via vector product
$-\,^{\pm}\!\vec{\omega}_\mu$ $\times$ $(\cdot)$. Denote by
$\,^{\pm}\!O_\mu$ = $\exp{(-\,^{\pm}\!\vec{\omega}_\mu
\times (\cdot))}$ thus obtained representation of
$\,^{\pm}\!\Omega_\mu$ = $\exp{(\,^{\pm}\!\omega_\mu)}$ in
the abstract space. The constraints $\pi^\alpha*\pi^\beta$
= 0 look as
\begin{equation}
\,^-\!\vec{\pi}^\alpha\cdot\,^-\!\vec{\pi}^\beta =
\,^+\!\vec{\pi}^\alpha\cdot\,^+\!\vec{\pi}^\beta .
\end{equation}
This defines $\,^-\!\vec{\pi}^\alpha$ in terms of
$\,^+\!\vec{\pi}^\alpha$ up to an overall SO(3,C) rotation
$U$ so that
\begin{equation}
\,^-\!\vec{\pi}^\alpha = U\vec{\pi}^\alpha ,
~~~\vec{\pi}^\alpha \equiv \,^+\!\vec{\pi}^\alpha .
\end{equation}
Substituting this into $\ref{d-mu}$ we find that $U$ can be
absorbed by $\,^-\!O_\mu$ so let us denote
\begin{equation}
G_\mu \equiv \,^-\!O_\mu U ,~~~O_\mu \equiv \,^+\!O_\mu .
\end{equation}
The Lagrangian on the constraint surface (kinetic term)
takes the form
\begin{equation}
L = \sum_{\alpha}{{1\over 4}\varepsilon^{ikl}\pi^\alpha_i
(\bar{O}_\alpha\dot{O}_\alpha - \bar{O}_0\dot{O}_0
+ \bar{G}_\alpha\dot{G}_\alpha - \bar{G}_0\dot{G}_0)_{kl}}
+ \sum_{\mu > \nu}{\dot{\psi}_{\mu\nu}\vec{\pi}^\mu\cdot
\vec{\pi}^\nu}.
\end{equation}
The Gaussian constraint splits into
\begin{eqnarray}
\sum_{\alpha}{(O_\alpha - O_0)\vec{\pi}^\alpha} & = & 0,\\
\sum_{\alpha}{(G_\alpha - G_0)\vec{\pi}^\alpha} & = & 0.
\end{eqnarray}
Analogously, for curvature matrix $R_{\mu\nu}$ denote by
$\,^{\pm}\!P_{\mu\nu}$ representations of $\,^{\pm}\!
R_{\mu\nu}$ ($R_{\mu\nu}$ = $\,^+\!R_{\mu\nu}$ $\,^-\!
R_{\mu\nu}$ = $\,^-\!R_{\mu\nu}$ $\,^+\!R_{\mu\nu}$) in the
abstract space. Consider for definiteness $R_{01}$ =
$\bar{\Omega}_{02}$ $\Omega_{03}$. Absorbing $U$ by $\,^-\!
P_{01}$ we find
\begin{eqnarray}
P_{01} & \equiv & \,^+\!P_{01} = \exp{(\,^+\!\varphi^2_0
\vec{\pi}^2\times (\cdot))}\bar{O}_2O_3\exp{(-\,^+\!
\varphi^3_0\vec{\pi}^3\times (\cdot))},\\
Q_{01} & \equiv & \bar{U}\,^-\!P_{01}U = \exp{(\,^-\!
\varphi^2_0\vec{\pi}^2\times (\cdot))}\bar{G}_2G_3\exp{
(-\,^-\!\varphi^3_0\vec{\pi}^3\times (\cdot))},\\
\,^{\pm}\!\varphi^\nu_\mu & \equiv & \varphi^\nu_\mu \pm
\,^*\!\varphi^\nu_\mu.
\end{eqnarray}
The corresponding contribution to the Hamiltonian reads
\begin{equation}
\label{n-alpha}
|n_{01}|\alpha_{01} = |\vec{n}_{01}|\arcsin{{n^i_{01}
\varepsilon_{ikl}\over 8|\vec{n}_{01}|}(P^{kl}_{01}\sqrt{1
+ {\rm tr}Q_{01}} + Q^{kl}_{01}\sqrt{1
+ {\rm tr}P_{01}})}
\end{equation}
where $\vec{n}_{01}$ = $u^2_0\vec{\pi}^3$ -
$u^3_0\vec{\pi}^2$ - $\vec{\pi}^2\times\vec{\pi}^3$.

Return now to the path integral. Integration of six
$\delta$-functions $\delta (\vec{\pi}^\alpha\cdot\vec{\pi}
^\beta - \,^-\!\vec{\pi}^\alpha\cdot\,^-\!\vec{\pi}^\beta)$
over $d^3\,^-\!\vec{\pi}^1$ $d^3\,^-\!\vec{\pi}^2$ $d^3
\,^-\!\vec{\pi}^3$ results in $V^{-2}{\cal D}^3U$ ($V^2$ =
$\vec{\pi}^1\times\vec{\pi}^2\cdot\vec{\pi}^3$). On the
other hand, since $U$ is absorbed by "-" components of
connection matrices and due to the invariance of the
measure ${\cal D}^6\Omega_\mu$ = ${\cal D}^3\,^+\!O_\mu
{\cal D}^3\,^-\!O_\mu$ = ${\cal D}^3O_\mu{\cal D}^3G_\mu$
the integral over ${\cal D}^3U$ decouples giving the volume
of the gauge subgroup. This leads to the following
replacement in the measure (\ref{d-mu}):
\begin{equation}
\prod_{\alpha \geq \beta}{\delta (\pi^\alpha*\pi^\beta )}
\prod_{\mu}{{\cal D}^6\Omega_\mu}
\prod_{\alpha}{d^6\pi^\alpha}\Rightarrow
{1\over \vec{\pi}^1\times\vec{\pi}^2\cdot\vec{\pi}^3}
\prod_{\alpha}{d^3\vec{\pi}^\alpha}
\prod_{\mu}{{\cal D}^3O_\mu{\cal D}^3G_\mu}
\end{equation}

The procedure of the derivation of the Ashtekar action
by the selfdual-antiselfdual splitting of the continuum GR
action in the HP form is known in detail (see, e.g.,
reviews \cite{Rom,Pel}). We find the following difference
of our discrete case from the continuum one: now there is
no the constraints like the so-called "reality conditions"
of the continuum GR which would relate $G_\mu$ and $O_\mu$.
Rather we have a theory with one set of area vectors
$\vec{\pi}^\mu$ but two sets of connections $G_\mu$ and
$O_\mu$. The situation is complicated by the fact that the
dependences of the Hamiltonian on $G_\mu$ and $O_\mu$ does
not split, see (\ref{n-alpha}). Nevertheless, this
procedure remains useful, for it allows to remove a part of
the gauge degrees of freedom in an invariant way.

\bigskip
This work was supported in part by the RFBR grant
No. 96-15-96317.

\end{document}